\documentclass[12pt]{article}

\usepackage{amssymb,amsmath,bm,bbm}
\usepackage{cite}
\usepackage{color}
\usepackage{graphicx}
\usepackage{slashed} 
\usepackage{hyperref}

\begin{document}
\begin{titlepage}
\vspace{0.4truecm}

\title{\Large\textbf{Enhancement of  dark matter relic density from \\
the late time dark matter conversions}}

\vspace{0.4truecm}

\author{ 
Ze-Peng Liu\footnote{Email: zpliu@itp.ac.cn}, 
Yue-Liang Wu\footnote{Email: ylwu@itp.ac.cn},  \  and 
Yu-Feng Zhou\footnote{Email: yfzhou@itp.ac.cn}\\ \\
%
\textit{Kavli Institute for Theoretical Physics China,}\\
\textit{Institute of Theoretical Physics,} \\
\textit{Chinese Academy of Sciences, Beijing, 100190, P.R. China }
}

\date{}
\maketitle
\begin{abstract}
\noindent
We demonstrate that if the dark matter (DM) in the Universe contains
multiple components, the possible interactions between the DM components may convert
the heavier DM components into the lighter ones. It is then possible that
the lightest DM component with an annihilation cross section significantly larger
than that of the typical weakly interacting massive particle (WIMP) can
obtain a relic density in agreement with the cosmological observations, due
to an enhancement of number density from the DM conversion process at late
time after the thermal decoupling, which may provide an alternative source of
boost factor relevant to  the positron and electron excesses reported by the recent
DM indirect search experiments.
\end{abstract}

\end{titlepage}

\section{Introduction}
In the recent years, a number of experiments such as PAMELA
\cite{Adriani:2008zr}, ATIC \cite{:2008zzr}, Fermi-LAT \cite{Abdo:2009zk} and
HESS \cite{Aharonian:2009ah} etc.  have reported excesses in the high energy
spectrum of cosmic-ray positrons and electrons over the backgrounds estimated
from the traditional astrophysics. Besides plausible astrophysical
explanations~\cite{Hooper:2008kg,Yuksel:2008rf,Profumo:2008ms}, the dark
matter (DM) annihilation or decay provides exciting alternative explanations
from particle physics.

If the DM particle is a thermal relic such as the weakly interacting massive
particle (WIMP), the thermally averaged product of its annihilation cross
section with the relative velocity at the time of thermal freeze out is typically
$\langle\sigma v\rangle_F\simeq 3\times
10^{-26}\mbox{cm}^{3}\mbox{s}^{-1}$. The positron or electron flux produced by
the DM annihilation can be parametrized as
\begin{eqnarray}
\Phi_{e} & = & B\overline{N}_{e}\frac{\rho_{0}^{2}\langle\sigma v\rangle_F}{m_D^{2}} ,
\end{eqnarray}
where $\rho_{0}\simeq0.3\mbox{ GeV}\cdot\mbox{cm}^{-3}$ is the smooth local DM
energy density estimated from astrophysics, $\overline{N}_{e}$ is the averaged
electron number produced per DM annihilation which depends on DM models and
propagation parameters, and $m_D$ is the mass of the DM particle. The boost
factor $B$ is defined as $B\equiv(\rho/\rho_{0})^{2}\langle\sigma
v\rangle/\langle\sigma v\rangle_F$ with $\rho$ the true local DM density and
$\langle \sigma v \rangle$ the  DM annihilation cross section multiplied 
by the relative velocity and averaged over the DM velocity distribution  today.
Both the PAMELA and Fermi-LAT results indicate that a large
boost factor is needed~\cite{Cholis:2008hb,Bergstrom:2009fa}. For a typical DM
mass of $\sim$1(1.6) TeV the required boost factor $B$ is $\sim 500(1000)$ for
DM annihilating directly into $\mu^+\mu^-$ and $\rho$ fixed to
$\rho_0$ ~\cite{Bergstrom:2009fa}. 

A large boost factor may arise from the non-uniformity of the DM distribution
in the DM halo. The N-body simulations show however that the local cumps of dark
matter density are unlikely to contribute to a large enough
$\rho/\rho_{0}$~\cite{Springel:2008by,Diemand:2008in}.
An other possibility of increasing the boost factor is that the DM
annihilation cross section may be velocity-dependent which grows at low
velocity. The DM annihilation cross section today may be much larger than that
at the time of thermal freeze out, and thus is not constrained by the DM relic
density. Some enhancement mechanisms have been proposed along this line, such
as the Sommerfeld enhancement~\cite{Sommerfeld,Hisano:2002fk,Hisano:2003ec,Cirelli:2007xd,ArkaniHamed:2008qn,Pospelov:2008jd,MarchRussell:2008tu,Iengo:2009ni,Cassel:2009wt} and the
resonance enhancement~\cite{Feldman:2008xs,Ibe:2008ye,Guo:2009aj} etc.

In some non-thermal DM scenarios, the number density of the DM particle can be
enhanced by the out of equilibrium decay of some heavier unstable particles if
the DM particle is among the decay products of the decaying
particle~\cite{Fairbairn:2008fb,Feldman:2009wv}.  The decay of the unstable
particle must take place at very late time. Otherwise the DM particles with
the enhanced number density will annihilate into the Standard Model (SM)
particles again, which washes out the effect of the enhancement. This requires
that the decay width of the unstable particle must be extremely small,
typically $10^{-17}$ GeV for the mass of the decaying particle around TeV
\cite{Fairbairn:2008fb}, which is much smaller than that of the typical weak
interaction.

In this work, we consider an alternative possibility for generating a 
boost factor, which does not require the velocity-dependent annihilation cross
section or the decay of unstable particles.  We show that in the scenarios of
interacting Milt-component DM, the interactions among the DM components may
convert the heavier DM components into the lighter ones, which is not
sensitive to the details of the conversion interaction.  If the interactions
are strong enough and the DM components are nearly degenerate in mass, the
conversion can enhance the number density of the lighter DM components at late
time after the thermal decoupling. Eventually, the whole DM today in the
Universe can be dominated by the lightest DM component with enhanced number
density, which corresponds to a large boost factor. The scenarios of
multi-component DM have been discuss previously in
Refs.~\cite{Boehm:2003ha,Hur:2007ur,Adibzadeh:2008pe,Feng:2008ya,Zurek:2008qg,Batell:2009vb,Profumo:2009tb,Zhang:2009dd,Gao:2010pg,Feldman:2010wy}.
Note however that the models with simply mixed non-interacting multi-component
DM cannot generate large boost factors.

This paper is organized as follows: in section \ref{evolution}, we first 
discuss the thermal evolution of the DM number densities in generic multi-component
DM models. We then give approximate analytic expressions as well as precise 
numerical calculations of the boost factor in a generic two-component DM model. In 
section \ref{model}, we consider a concrete model containing two fermionic DM particles with
extra $U(1)$ gauge interactions in the hidden sector. The conclusions are given in 
section \ref{conclusion}.


\section{Thermal evolution of the interacting  multi-component DM}\label{evolution}
Let us consider a generic model in which the whole cold DM contains $N$
components $\chi_{i}\ (i=1,\dots,N)$, with masses $m_{i}$ and internal degrees
of freedom $g_{i}$ respectively. The DM components are labeled such that
$m_{i}<m_{j}$ for $i<j$, thus $\chi_{1}$ is the lightest DM particle. We are
interested in the case that $\chi_{i}$ are nearly degenerate in mass, namely
the relative mass differences between $\chi_i$ and $\chi_1$ satisfy
$\varepsilon_{i}\equiv(m_{i}-m_{1})/m_{1}\ll1$. In this case, we shall show
that the interactions between the DM components lead to the DM conversion. 
The situation is analogous to the neutral meson mixing and neutrino oscillations
in particle physics. They all occur at small mass differences.
The thermal evolution of the DM number density normalized to the entropy density
$Y_{i}\equiv n_{i}/s$ with respect to the rescaled temperature $x\equiv
m_{1}/T$ is govern by the following Boltzmann equation
\begin{eqnarray}
&&\frac{dY_{i}(x)}{dx} =
-\frac{\lambda}{x^{2}}
\left[
\langle\sigma_{i}v\rangle(Y_{i}^{2}-Y_{ieq}^{2})
-\sum_{j}\langle\sigma_{ij}v\rangle(Y_{i}^{2}-r_{ij}^{2}Y_{j}^{2})
\right] , 
\label{Boltzmann-eq}
\end{eqnarray}
where $\lambda\equiv x s/H(T)$ is a
combination of $x$, the entropy density $s$ and the Hubble parameter $H(T)$ as a function
of temperature $T$. 
$Y_{ieq}\simeq (g_{i}/s)[m_{i}T/(2\pi)]^{3/2}\exp(-\varepsilon_i x)$ is
the equilibrium number density normalized to entropy density for
non-relativistic particles.  $\langle\sigma_{i}v\rangle$ are the thermally
averaged cross sections multiplied by the DM relative velocity for the process
$\chi_{i}\chi_{i}\to XX'$ with $XX'$ standing for the light SM particles which are
in thermal equilibrium, and $\langle\sigma_{ij}v\rangle$ are the ones for the DM
conversion process $\chi_{i}\chi_{i}\to\chi_{j}\chi_{j}$.
The quantity 
\begin{equation}\label{ratio-r}
r_{ij}(x)\equiv \frac{Y_{ieq}(x)}{Y_{jeq}(x)}
=\left(\frac{g_i}{g_j} \right)
\left(\frac{m_i}{m_j} \right)^{3/2}
\exp[-(\epsilon_i-\epsilon_j)x]
\end{equation}
is the ratio between the two equilibrium number density functions, In writing
down Eq. (\ref{Boltzmann-eq}) we have assumed kinetic equilibrium.  The first
term in the r.h.s. of Eq.(\ref{Boltzmann-eq}) describes the change of number
density of $\chi_i$ due to the annihilation into the SM particles, while the
second term describes the change due to the conversion to other DM particles.

In the case that the cross section of the conversion process $\langle
\sigma_{ij}v \rangle$ is large enough, the DM particle $\chi_i$ can be kept in
thermal equilibrium with $\chi_j$ for a long time after both $\chi_i$ and
$\chi_j$ have decoupled from the thermal equilibrium with the SM particles.
In this case, the number densities of $\chi_{i,j}$ satisfy a simple relation
\begin{equation}\label{ratio}
\frac{Y_{i}(x)}{Y_{j}(x)}\approx \frac{Y_{ieq}(x)}{Y_{jeq}(x)}=r_{ij}(x) .
\end{equation}   
We  emphasize that even when $\chi_{i}$ is in equilibrium with
$\chi_j$ the ratio of the number density $Y_{i}(x)/Y_{j}(x)$ can be quite
different from unity and can vary with temperature. For instance, if $g_i\gg
g_j$ and $0< (\epsilon_i-\epsilon_j) \ll 1$, from Eq. (\ref{ratio-r}) and
(\ref{ratio}) one obtains $Y_{i}(x)\gg Y_{j}(x)$ at the early time when
$(\epsilon_i-\epsilon_j)x \ll 1$. However, at the late time when
$(\epsilon_i-\epsilon_j)x \gg 1$, one gets $Y_{i}(x)\ll Y_{j}(x)$, which is
due to the Boltzmann suppression factor $\exp[-(\epsilon_i-\epsilon_j)x]$ in the
expression of $r_{ij}$. Thus the heavier particles can be gradually converted into
lighter ones through this temperature-dependent equilibrium between $\chi_i$
and $\chi_j$.

Since all the DM components $\chi_i$ are stable, in general the co-annihilation
process $\chi_{i}\chi_{j}\to XX'$ are not allowed as the crossing process
$\chi_i\to\chi_{j} XX'$ corresponds to the decay of $\chi_i$. Furthermore,
unlike the case of co-annihilation, $\chi_i$ and $\chi_j$ may not necessarily
share the same quantum numbers.

An interesting limit to consider is that the rates of DM conversion  are large
compared with that of the individual DM annihilation into the SM particles,
i.e.  $\langle\sigma_{ij}v\rangle \gtrsim \langle\sigma_{i}v\rangle$.  In this
limit, after both the DM particles have decoupled from the thermal equilibrium
with the SM particles, which take place at a typical temperature $x=x_{dec}\approx25$, the strong
interactions of conversion will maintain an equilibrium between $\chi_i$ and $\chi_j$ 
for a long time
until the rate of the conversion cannot compete with the expansion rate of the
Universe.  Making use of Eq. (\ref{ratio}),  the evolution of the total density
$Y(x)\equiv\sum_{i=1}^{N}Y_{i}(x)$ can be written as
\begin{eqnarray}\label{evolution-sum}
\frac{dY}{dx} & = & 
-\frac{\lambda}{x^{2}}\langle\sigma_{eff}v\rangle\left(Y^{2}-Y_{eq}^{2}\right) ,
\end{eqnarray}
where $\langle\sigma_{eff}v\rangle$ is the effective thermally averaged 
product of DM annihilation cross section and the relative velocity which 
can be written as
\begin{eqnarray}
\langle\sigma_{eff}v\rangle
=
\frac{\sum_{i=1}^{N}w_{i}g_{i}^{2}(1+\varepsilon_{i})^{3}\exp(-2\varepsilon_{i}x)}{g_{eff}^{2}} \langle\sigma_{1}v\rangle ,
\end{eqnarray}
where  $w_{i}\equiv\langle\sigma_{i}v\rangle/\langle\sigma_{1}v\rangle$
is the  annihilation cross section relative to that of the lightest one.
The total equilibrium number density can be written as
\begin{equation}
Y_{eq}\equiv\sum_{i=1}^{N}Y_{ieq}(x)\approx
g_{eff}\left(
  \frac{m_{1}T}{2\pi}
\right)^{3/2} \exp(-x)  ,
\end{equation}
with  effective degrees of
freedom $g_{eff}=\sum_{i}g_{i}(1+\varepsilon_{i})^{3/2}\exp(-\varepsilon_{i}x)$.  
Note that the conversion terms do not show up explicitly in
Eq. (\ref{evolution-sum}).  Through the conversion processes
$\chi_i\chi_{i}\to \chi_j\chi_{j}$ the slightly heavier components will be
converted into the lighter ones, because the factor $r_{ij}(x)$ is
proportional to $\exp[-(m_i-m_j)/T]$ which suppresses the density of the
heavier components at lower temperature.  If the conversion cross section is
large enough, most of the DM components will be converted into the lightest
$\chi_{1}$ before the interaction of conversion decouples, which may result in
a large enhancement of the relic density of $\chi_1$ and leads to a large
boost factor.

As an example, let us  consider a generic DM model with only two
components. For relatively large conversion cross section $u \equiv \langle
\sigma_{21}v \rangle /\langle \sigma_1 v\rangle \gtrsim 1$, The effective
total cross section is given by
\begin{equation}
\langle\sigma_{eff}v\rangle=\frac{1+w g^{2}\exp(-2\varepsilon
x)}{[1+g\exp(-\varepsilon x)]^{2}}\langle\sigma_{1}v\rangle ,
\end{equation}
where $w\equiv w_2 $, $g\equiv g_{2}/g_{1}$ and
$\varepsilon\equiv\varepsilon_{2}$.  Because of the $x$-dependence in
$\langle\sigma_{eff}v\rangle$, the thermal evolution of $Y(x)$ differs
significantly from that of the standard WIMP. In the case that $\chi_2$ has
large degrees of freedom but a small annihilation cross section,
namely $g\gg1$, $w\ll1$ and $wg^{2}\ll1$, the thermal evolution of the total
density $Y$ can be approximated by
\begin{eqnarray}
\frac{dY}{dx} & \approx & -\frac{\lambda}{x^{2}}\frac{1}{[1+g\exp(-\varepsilon x)]^{2}}\langle\sigma_{1}v\rangle\left(Y^{2}-Y_{eq}^{2}\right) ,
\label{2dm-eq}
\end{eqnarray}
the thermal evolution of the total number density can be roughly divided into four stages:  
i) At high temperature region where $3 \lesssim x\ll x_{dec}$, both the DM
components are in thermal equilibrium with the SM particles.  $Y_{i}(x)$ must
closely track $Y_{ieq}(x)$ which decrease exponentially as $x$ increases.
However,  since $g\gg 1$ and
$\epsilon \ll 1$, the number density of $\chi_2$ is much higher than
that of $\chi_1$, i.e. $Y_2(x) \gg Y_1(x)$.
ii) When the temperature goes down and $x$ is close to the decoupling point
$x_{dec}$, both the DM components start to decouple from the thermal
equilibrium. In the region $x_{dec}\lesssim x\ll 1/\varepsilon$,
$\langle\sigma_{eff}v\rangle$ is nearly a constant and
$\langle\sigma_{eff}v\rangle\approx\langle\sigma_{1}v\rangle/(1+g)^{2}\ll\langle\sigma_{1}v\rangle$,
the total density $Y(x)$ behaves just like that of an ordinary WIMP which
converges quickly to $Y(x)\approx x_{dec}/(\lambda \langle \sigma_1 v
\rangle)$.
iii) As $x$ continues growing, the suppression factor $\exp(-\varepsilon x)$
in $\langle\sigma_{eff}v\rangle$ becomes relevant. The value of
$\langle\sigma_{eff}v\rangle$ grows rapidly especially after $x$ reaches the
point $\varepsilon x\approx\mathcal{O}(1)$, which leads to the further
reduction of $Y(x)$. In this stage, although both $\chi_{1,2}$ have decoupled
from the thermal equilibrium with the SM particles. The strong conversion
interaction $\chi_2\chi_2 \leftrightarrow \chi_1\chi_1$ maintains an
equilibrium between the two DM components. According to Eq. (\ref{ratio}), the
relative number density $Y_2(x)/Y_1(x)$ decreases with $x$ increasing, which
corresponds to the conversion from the heavier DM component into the lighter
one.  At the point $x_c= (1/\varepsilon)\ln g$ one has
$Y_2(x)\approx Y_1(x)$. 
For the region $x>x_{dec}$ and $x$ is not close to
$x_c$, because of $Y_{eq}(x)\ll Y(x)$ and $g\exp(-\varepsilon x)\gg 1$,
the Eq.~(\ref{2dm-eq}) can be analytically integrated out, using the expression
$I(x)=\int x^{-2} \exp(x)dx=\mbox{Ei}(x)-\exp(x)/x$ where $\mbox{Ei}(x)$ is
the exponential integral function. The integral has an asymptotic form of
$I(x)\approx \exp(x)/x^{2}$ for $x\gg1$. Thus $Y(x)$ in this region can be approximated 
by
\begin{equation}
Y(x)\approx \frac{g^2 x_{dec}}{\lambda \langle \sigma_1 v\rangle }
\left[ 1+
\left(\frac{ x_{dec}}{x}\right)\frac{\exp(2\varepsilon x )}{2\varepsilon x}
\right]^{-1}  .
\label{approx-solution}
\end{equation}
iv) When $x$ becomes very large $\varepsilon x\gg\mathcal{O}(1)$ ,
$\langle\sigma_{eff}v\rangle$ quickly approaches $\langle\sigma_{1}v\rangle$,
and becomes independent of $x$ again. The evolution of $Y(x)$ in this region can be
obtained by a simple integration as it was done in the stage ii). The solution of $Y(x)$
shows a second decoupling.  Finally when the conversion rate cannot compete
with the expansion rate of the Universe at some point $x_F$ corresponding to $
sY_2\langle \sigma_{21} v\rangle/H \approx 1$, both $Y_1(x)$ and $Y_2(x)$ remain
unchanged as relics. The whole DM can be dominated by $\chi_{1}$ if
the conversion is efficient enough.

By matching the analytic solutions of $Y(x)$ in different regions near the points
$x_{dec}$ and $x_c$, and requiring that the final total relic density is
equivalent to the observed $\Omega_{CDM} h^2\approx 0.11$, we obtain the following
approximate expression of the boost factor
\begin{equation}
B\approx g^2 \left[1+ \left(\frac{x_{dec}}{x_c}\right) \left(\frac{\exp(2\varepsilon x_c)}{2\varepsilon x_c}+g^2 \right) \right]^{-1} .
\label{boostFacEq}
\end{equation}
As expected, the enhancement essentially comes from the conversion of the
degrees of freedom. Thus the maximum enhancement is $g^2$. The two 
terms in the r.h.s of the above equation correspond to the reduction of $Y(x)$
during the late time conversion stages.  For large enough $g$, the boost
factor can be approximated by $B\approx g^2/(1+\varepsilon g^2 x_{dec}/\ln g
)$. In order to have a large boost factor, a small $\varepsilon \ll \ln g/(g^2
x_{dec})$ is also required. As shown in Eq. (\ref{boostFacEq}) the boost factor is
not sensitive to the exact values of the cross sections as long as the
conditions $w \ll 1$ and $u\gg 1$ are satisfied.

\begin{figure}[htb]
\begin{center}
\includegraphics[width=0.65\columnwidth]{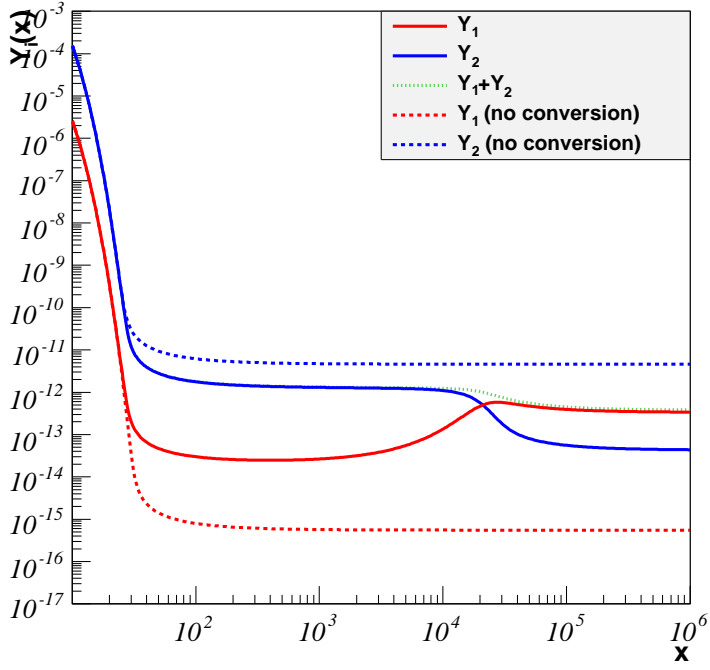}
\end{center}
\caption{
Thermal evolution of the number densities $Y_1(x)$ (red solid) and $Y_2(x)$ (blue solid) 
with respect to $x$. The  solid (dashed) curves correspond  to the case with (without) DM 
conversions. The green dotted curve  corresponds to the sum of $Y_1$ and $Y_2$,  
for parameters $g=60$, $m_1=1$TeV,
 $\varepsilon =2\times 10^{-4}$, $w=10^{-4}$ and $u=10$ respectively.
 \label{fig:Time-evolution}
}

\end{figure}

We numerically calculate the thermal evolution of $Y_i(x)$ and the boost
factor without using approximations for a generic two-component DM model. The
results for $w=10^{-4}$, $u=10$ and $\varepsilon=2\times 10^{-4}$ is shown in
Fig.~\ref{fig:Time-evolution}.  The value of $\langle\sigma_{2}v\rangle$ is
adjusted such that the final total DM relic abundance is always equal to the
observed value $\Omega_{CDM}h^{2}$.  The mass of the light DM particle is set
to $m_{1}=1$ TeV. For an illustration the ratio between the internal degrees
of freedom is set to be large $g=60$.
From the figure, the four stages of the thermal evolution of $Y(x)$ as well as
the crossing point can be clearly seen.  The crossing point at $x=x_c \approx
2\times 10^{-4}$ indicates the time when the number density of $\chi_{1}$
start to surpass that of $\chi_{2}$ and eventually dominant the whole DM relic
density. In this parameter set a large boost factor
$B\approx\langle\sigma_{1}v\rangle/\langle\sigma v\rangle_F\approx 585 $ is
obtained which is in a remarkable agreement with Eq. (\ref{boostFacEq}) with
error less than $\sim 5\%$.  For a comparison, in
Fig. \ref{fig:Time-evolution} we also show the cases without conversions.

\begin{figure}[htb]
\includegraphics[width=0.49\columnwidth]{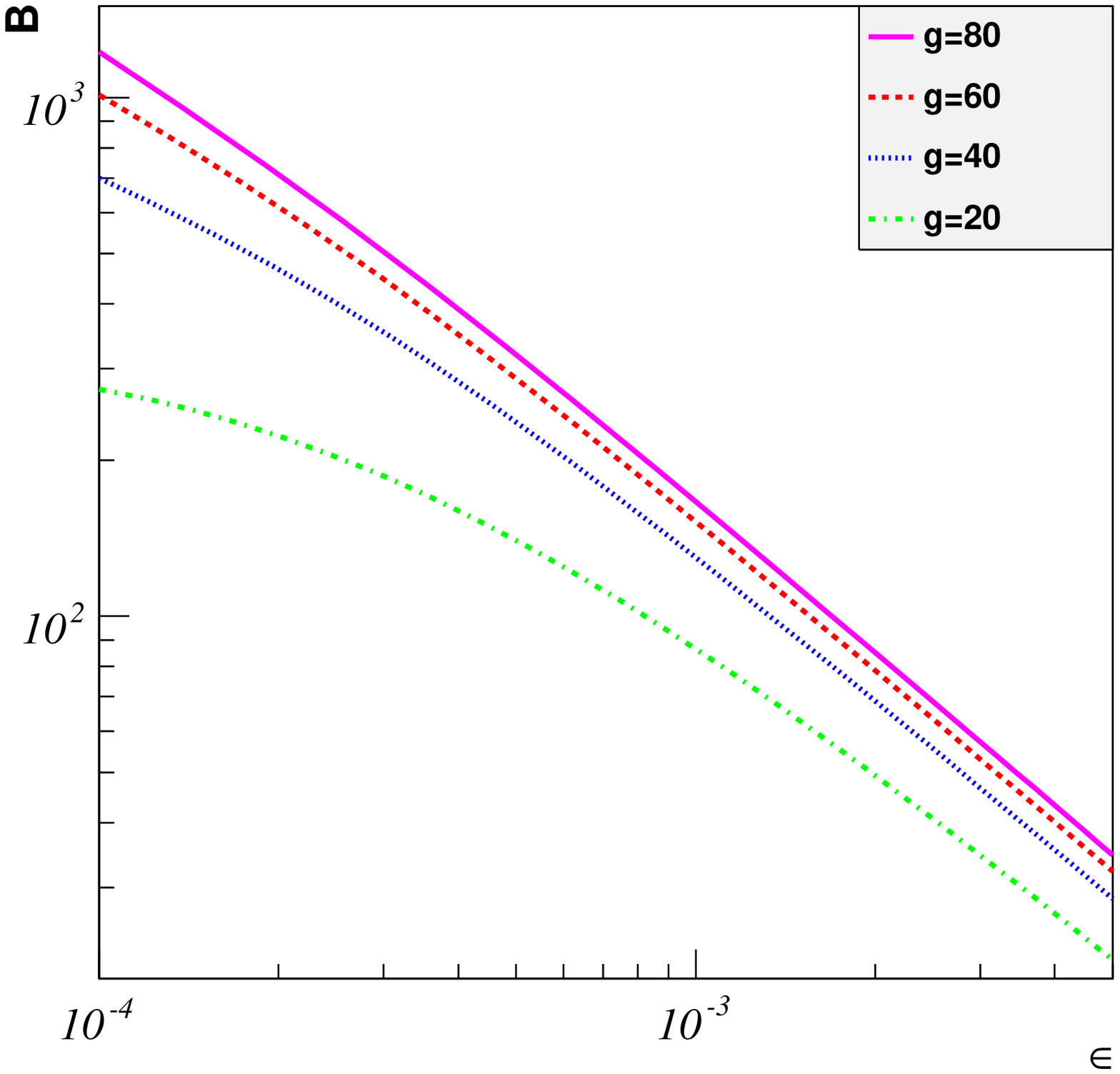} \includegraphics[width=0.49\columnwidth]{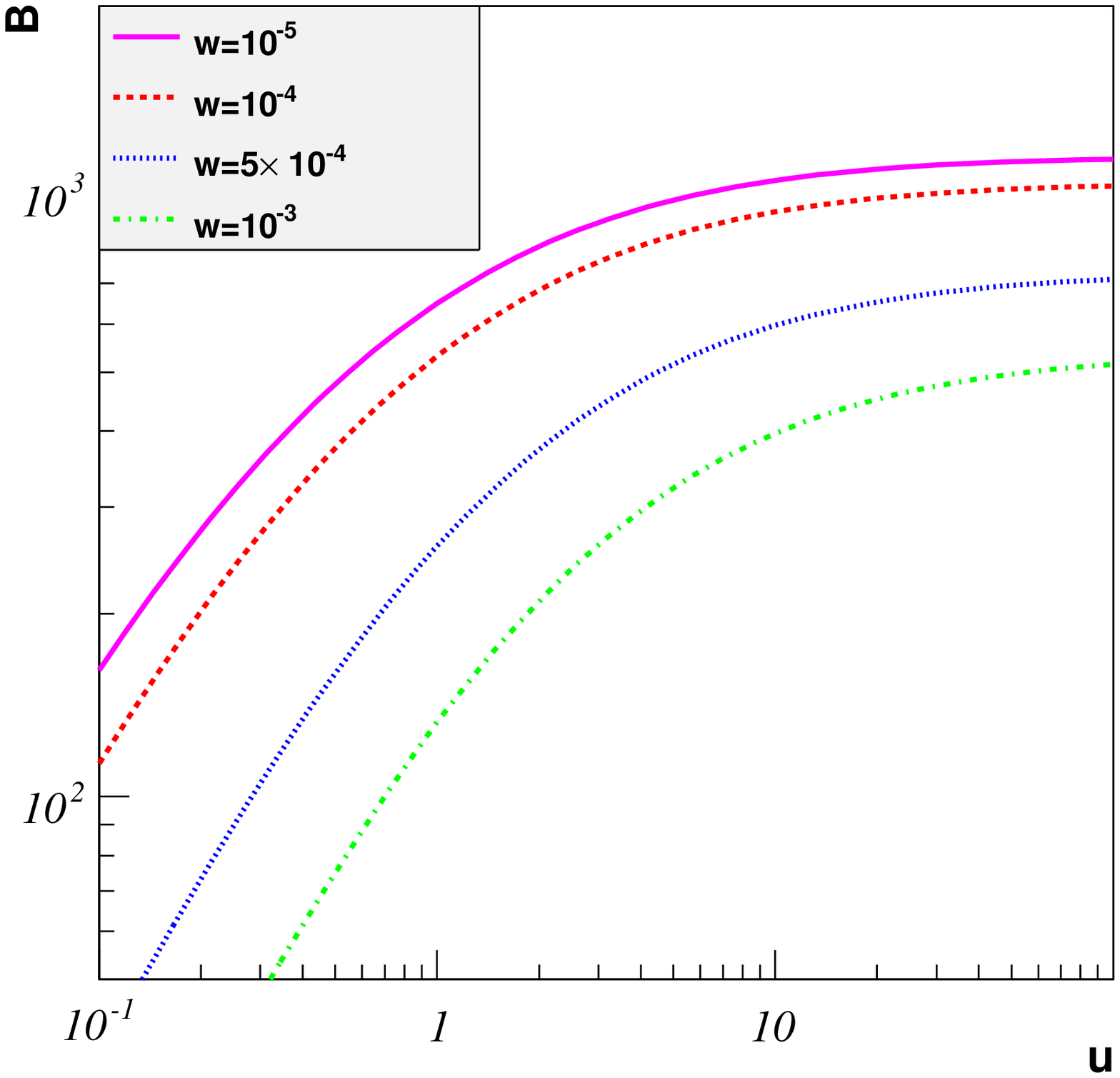}
\caption{
Left) boost factor $B$ as a function of the relative mass difference $\varepsilon$ for
different relative degrees of freedom $g$=80 (solid), 60 (dashed), 40 (dotted)
and 20 (dot-dashed)  respectively,  for $w=10^{-4}$ and $u=10^{2}$;
Right) boost factor as a function of the relative conversion cross section
$u$. Four curves correspond to $w=10^{-5}$
(solid), $10^{-4}$ (dashed), $5\times 10^{-4}$ (dotted) and $10^{-3}$ (dot-dashed) respectively,
for parameters   $g=60$, $m_{1}=1$ TeV and  $\varepsilon= 1\times 10^{-4}$ respectively.
\label{fig:boost-factor}}
\end{figure}

In Fig. \ref{fig:boost-factor} (left), we show how the boost factor $B$ varies
with the mass difference $\varepsilon$ for different relative internal degrees
of freedom $g$.  In general, $B$ becomes larger for smaller $\varepsilon$ and
larger $g$. For $\varepsilon=10^{-4}$ and $g=60$, the boost factor can reach
$B \sim 10^{3}$. For a much smaller $g=20$ and a larger $\varepsilon=8\times
10^{-4}$, the boost factor can still reach $\mathcal{O}(100)$.
The dependence of $B$ on the cross sections $u$ and $w$ is
shown in Fig. \ref{fig:boost-factor} (right).  A small $w$ and large $u$ lead
to the increasing of $B$. However, for very small $w \lesssim 10^{-4}$ and
vary large $u \gtrsim 100$, the value of $B$ becomes insensitive to the exact
values of $w$ and $u$, which is also in agreement with the approximate solution
given in Eq.~(\ref{boostFacEq}).

\section{A simple model with DM conversion}\label{model}

For models with multiple DM components, it is nature that there exists interactions among
the DM components which may lead to the conversions among  them.  In this work we
consider a simple interacting two-component DM model by adding to the standard
model (SM) with two SM gauge singlet fermionic DM particles $\chi_{1,2}$. The
particles $\chi_{1,2}$ are charged under a local $U(1)$ symmetry which is
broken spontaneously by the vacuum expectation value (VEV) of a scalar field
$\phi$ through the Higgs mechanism.  The corresponding massive gauge boson is
denoted by $A$ which may cause the interaction $\bar{\chi_2}\chi_2
\leftrightarrow \bar{\chi_1}\chi_1$.  The stability of $\chi_{1,2}$ is
protected by two different global $U(1)$ number symmetries.
%
An SM gauge singlet pseudo-scalar $\eta$ is introduced as a messenger field
which couples to both the dark sector and the SM sector.
%
In order to have the leptophilic nature of DM annihilation, we also introduce
an SM $SU(2)_{L}$ triplet field $\Delta$ with the SM quantum number $(1,3,1)$
and flavor contents $\Delta=(\delta^{++},\delta^{+},\delta^{0})$.  The triplet
carries the quantum number $B-L$=2 such that it can couple to the SM
left-handed leptons $\ell_L$ through Yukawa interactions $\bar{\ell}^c_L
\Delta \ell_L$, but cannot couple to quarks directly.  The VEV of the triplet
has to be very small around eV scale, which is required by the smallness of
the neutrino masses. As a consequence, the couplings between one triplet and
two SM gauge bosons such as $\delta^{\pm\pm}W^{\mp}W^{\mp}$,
$\delta^{\pm}W^{\mp}Z^0$ and $\delta^{0}Z^0Z^0$ are strongly suppressed as
they are all proportional to the VEV of the triplet, which makes it difficult
for the triplet to decay even indirectly into quarks through SM gauge bosons
\cite{Gogoladze:2009gi,Guo:2010vy}. If $\eta$ has a stronger coupling to
$\Delta$ than that to the SM Higgs boson $H$ and $\phi$ then the annihilation
products of the dark matter particles $\chi_{1,2}$ will be mostly leptons.

%
The full Lagrangian of the model can be written as $\mathcal{L}=\mathcal{L}_{SM}+\mathcal{L}_1$
The new interactions in $\mathcal{L}_1$ which are relevant to the DM annihilation and conversion are 
given by
\begin{eqnarray}\label{lagrangian}
\mathcal{L}_1 
&\supset &
\bar{\chi}_i (i\slashed D-m_i )\chi_i
+(D_\mu \phi)^\dagger (D^\mu \phi)- m_\phi^2 \phi^\dagger\phi  
\nonumber\\
&&+\frac{1}{2}\partial_\mu\eta\partial^\mu\eta-\frac{1}{2} m_\eta^2 \eta^2 -y_i \bar{\chi}_i i\gamma_5 \eta \chi_i -y_\ell \bar{\ell}^c_L \Delta \ell_L+\mbox{h.c}
\nonumber\\
&& -(\mu \eta+\xi\eta^2 )
\left[
  \mbox{Tr}(\Delta^\dagger \Delta )+\kappa (H^\dagger H)+\zeta (\phi^\dagger\phi)
\right],  \;\; (i=1,2)
\end{eqnarray}
with $D_\mu =\partial_\mu +i g_A A_\mu$ and $g_A$ standing for the gauge
coupling constant.  Note that $\phi$ and $\eta$ do not directly couple to the
SM fermions. 
After the spontaneous symmetry breaking in $V(\phi)$, 
the scalar $\phi$ obtains a nonzero VEV $\langle \phi \rangle=v_\phi/\sqrt{2}$  
which generates the mass  of  the gauge boson  $m_A=g_A v_\phi$. 
At the tree level, the three components of the triplet $\delta^{++},\delta^{+}$
and $\delta^0$ are degenerate in mass, i.e.  
$m_{\delta^{++}}=m_{\delta^{+}}=m_{\delta^{+}}\equiv m_\Delta$. 

After the spontaneous symmetry breaking in the scalar sectors, the fields
$\Delta$, $H$ and $\phi$ obtain nonzero VEVs, which also generates a linear
term in $\eta$ through the last term of Eq. (\ref{lagrangian}).  The linear
term in $\eta$ in turn leads to a nonzero VEV of $\eta$, i.e., $\langle \eta
\rangle=v_\eta\neq 0$, which will give corrections to the masses of $\chi_i$
and may enlarge the mass difference between $\chi_1$ and $\chi_2$. This
problem can be avoided by using the above mentioned assumption that $\eta$ has
a much stronger coupling to $\Delta$ than that to $H$ and $\phi$, which
requires that $\kappa, \zeta \ll 1$. The VEV of $\eta$ is proportional to the
ratio between the linear and quadratic terms in $\eta$, and can be estimated
as $v_\eta \approx -\mu( \kappa v_H^2 + \zeta v_\phi^2)/(2(m_\eta^2+\kappa
v_H^2 + \zeta v_\phi^2))$. Since the VEV of the triplet $\Delta$ is extremely
small and $v_H\approx \mathcal{O}(10^2)\mbox{GeV}$, if $\mu$, $m_\eta$, and
$v_\phi$ are all around TeV scale, for $\kappa \lesssim \mathcal{O}(10^{-2})$
and $\zeta \lesssim\mathcal{O}(10^{-4})$ the VEV of $\eta$ is $v_\eta \lesssim
\mathcal{O}(10^{-4})\mbox{TeV}$ which is small enough to avoid breaking the
degeneracy in the masses of $\chi_{1,2}$.

We assume that $\chi_2$ has large internal degrees of freedom relative to that
of $\chi_1$, i.e., $g_2\gg g_1$, which can be realized if $\chi_2$ belongs to
a multiplet of the product of some global nonabelian groups.  For instance
$g_2=4\tilde{g}_2$ with $\tilde{g}_2=$16, 8, and 4 if it belongs to the spinor
representation of a single group of $SO(8)$, $SO(6)$ and $SO(4)$ respectively.
When $\chi_2$ belongs to a representation of the product of these groups, its
internal degrees of freedom can be very large.

At the early time when the temperature of the Universe is high enough, the
triplet $\Delta$ can be kept in thermal equilibrium with SM particles through
the SM gauge interactions.  The DM particles $\chi_i$ can reach thermal
equilibrium by annihilating into the triplet through the intermediate particle
$\eta$.  
The annihilation
$\bar{\chi}_2\chi_2\to\eta^*\to \delta^{\pm\pm}\delta^{\mp\mp},
\delta^{\pm}\delta^{\mp}, \delta^{0}\delta^{0*} $ is an $s$-wave process which
is dominant contribution . The cross section before averaging over the relative velocity $v$
is given by
\begin{equation}
\sigma_{i}v
=\frac{N_f y_{i}^{2}\mu^{2}}{16\pi g_{i} (s-m_{\eta}^{2} )^{2}}\sqrt{1-\frac{4m_{\Delta}^{2}}{s}} ,
\end{equation}
where $N_f=3$ is the number of final states, $m_\eta$ is the mass of $\eta$ and 
$s$ is the square of the total energy in the center of mass frame.  For
$s$-wave annihilation we use the approximation that the thermally averaged 
cross section is the same as the one before the average, i.e.,
$\langle \sigma v \rangle \simeq \sigma v$. From the above equation the ratio of the two annihilation cross
sections is $w=(y_2/y_1)^2(g_1/g_2)$.  It is easy to get a very small $w$ provided that
$y_2 \ll y_1$ and $g_1 \ll g_2$. In order to have a large enough $\langle \sigma_1 v \rangle \gg \langle \sigma v\rangle_F$ 
the product of the coupling constants $y_1 \mu$ must be large enough, 
or the squared mass  of $\eta$ is close to $s$.
The cross section of the conversion process $\bar{\chi}_2\chi_2\to A^* \to
\bar{\chi}_1\chi_1$ is given by
\begin{eqnarray}
\sigma_{12} v
=\frac{3 g_A^4 m^2_1 }{2\pi (s-m_A^2)^2}
\left( \frac{g_1}{g_2} \right)
\sqrt{1-\frac{4m_{1}^{2}}{s}} .
\end{eqnarray}
The cross section is suppress by $g_1/g_2$ and also the phase space factor
$\sqrt{1-4m_1^2/s}$ when $s$ is close to $4m_2^2$ at the vary late time of the
thermal evolution. However, the cross section be greatly enhanced if $m_A$ is
close to a resonance when the relation $s \simeq m_A^2$ is satisfied.  In the
numerical calculations, we find that for the following selected  parameters:
$m_1=1$TeV, $\epsilon=1\times 10^{-4}$, $g_1=1$, $g_2=60$, $m_{\Delta}=500$
GeV, $m_\eta=1.5$ TeV, $m_A=2.02$ TeV, $y_1=3$, $y_2=0.07$, $\mu/m_1=3$, and
$g_A=2.5$, the following ratio of the cross section can be obtained
$$
w\simeq 1\times 10^{-5}, \ u\simeq 0.5, \
\mbox{and} \  \langle \sigma_1 v \rangle /\langle \sigma v\rangle_F \simeq 500 .
$$
In this parameter set the relative mass difference between $m_A$ and $2m_2$ is around
$1\%$. From  Fig. \ref{fig:boost-factor}, one can see that the 
corresponding boost factor is $B\sim 500$, which is large enough to account for the
PAMELA data for the dark matter mass around TeV.  

\section{Discussions and Conclusions}\label{conclusion} 
The mechanism proposed here does not require velocity-dependent annihilation
cross sections which is essential to the Sommerfeld enhancement.  There exists
stringent constraints from astrophysical observations if the DM annihilation
cross section scales with velocity as $1/v$ or $1/v^2$ and saturates at very low
velocity. Those constraints involves the bound on the $\mu$-type distortion of
CMB spectrum~\cite{Zavala:2009mi,Hannestad:2010zt,Finkbeiner:2010sm} and the
bounds on diffuse gamma-rays from the cold structures which have lower
velocity dispersion than that in the solar neighborhood in which $v \sim
10^{-3}$.  For instance, in the subhalos the average velocity can be as low as
$v\sim 10^{-5}$~\cite{Lattanzi:2008qa}, and the DM velocity in the protohalos
can be even lower $v\sim 10^{-8}$~\cite{Kamionkowski:2008gj} .
 If the enhancement is insensitive to the velocity, those apstrophysical bounds can be
relaxed significantly.
Furthermore, unlike the Sommerfeld enhancement, no attractive long-range 
force between the DM particles is involved. The existence of such a long-range force 
can change the halo shape and is constrained by observations~\cite{Feng:2009mn,Feng:2009hw,Feng:2010zp}. The  boost factor from DM conversion is free from this type of 
constraint as well.

In summary, We have considered an alternative mechanism for obtaining
boost factors from DM conversions  which does not require the
velocity-dependent annihilation cross section or the decay of unstable
particles.  We have shown that if the whole DM is composed of multiple
components, the relic density of each DM component may not necessarily be
inversely proportional to its own annihilation cross section. We demonstrate
the possibility that the number density of the lightest DM component with an
annihilation cross section much larger than $\langle\sigma v\rangle_F$ can get
enhanced in late time through DM conversation processes, and finally dominates
the whole relic abundance, which corresponds to a  boost factor needed to
explain the excesses in cosmic-ray positron and electrons reported by the
recent experiments.

\section*{Acknowledgments}
This work is supported in part by the National Basic Research Program
of China (973 Program) under Grants No. 2010CB833000; the National
Nature Science Foundation of China (NSFC) under Grants No. 10975170,
No. 10821504 and No. 10905084; and the Project of Knowledge Innovation
Program (PKIP) of the Chinese Academy of Science.

\begin{thebibliography}{10}

\bibitem{Adriani:2008zr}
{\bf PAMELA} Collaboration, O.~Adriani {\em et.~al.}, {\it {An anomalous
  positron abundance in cosmic rays with energies 1.5-100 GeV}},  {\em Nature}
  {\bf 458} (2009) 607--609, [\href{http://xxx.lanl.gov/abs/0810.4995}{{\tt
  arXiv:0810.4995}}].

\bibitem{:2008zzr}
J.~Chang {\em et.~al.}, {\it {An excess of cosmic ray electrons at energies of
  300-800 GeV}},  {\em Nature} {\bf 456} (2008) 362--365.

\bibitem{Abdo:2009zk}
{\bf The Fermi LAT} Collaboration, A.~A. Abdo {\em et.~al.}, {\it {Measurement
  of the Cosmic Ray e+ plus e- spectrum from 20 GeV to 1 TeV with the Fermi
  Large Area Telescope}},  {\em Phys. Rev. Lett.} {\bf 102} (2009) 181101,
  [\href{http://xxx.lanl.gov/abs/0905.0025}{{\tt arXiv:0905.0025}}].

\bibitem{Aharonian:2009ah}
{\bf H.E.S.S.} Collaboration, F.~Aharonian {\em et.~al.}, {\it {Probing the
  ATIC peak in the cosmic-ray electron spectrum with H.E.S.S}},  {\em Astron.
  Astrophys.} {\bf 508} (2009) 561,
  [\href{http://xxx.lanl.gov/abs/0905.0105}{{\tt arXiv:0905.0105}}].

\bibitem{Hooper:2008kg}
D.~Hooper, P.~Blasi, and P.~D. Serpico, {\it {Pulsars as the Sources of High
  Energy Cosmic Ray Positrons}},  {\em JCAP} {\bf 0901} (2009) 025,
  [\href{http://xxx.lanl.gov/abs/0810.1527}{{\tt arXiv:0810.1527}}].

\bibitem{Yuksel:2008rf}
H.~Yuksel, M.~D. Kistler, and T.~Stanev, {\it {TeV Gamma Rays from Geminga and
  the Origin of the GeV Positron Excess}},  {\em Phys. Rev. Lett.} {\bf 103}
  (2009) 051101, [\href{http://xxx.lanl.gov/abs/0810.2784}{{\tt
  arXiv:0810.2784}}].

\bibitem{Profumo:2008ms}
S.~Profumo, {\it {Dissecting Pamela (and ATIC) with Occam's Razor: existing,
  well-known Pulsars naturally account for the 'anomalous' Cosmic-Ray Electron
  and Positron Data}},  \href{http://xxx.lanl.gov/abs/0812.4457}{{\tt
  arXiv:0812.4457}}.

\bibitem{Cholis:2008hb}
I.~Cholis, L.~Goodenough, D.~Hooper, M.~Simet, and N.~Weiner, {\it {High Energy
  Positrons From Annihilating Dark Matter}},  {\em Phys. Rev.} {\bf D80} (2009)
  123511, [\href{http://xxx.lanl.gov/abs/0809.1683}{{\tt arXiv:0809.1683}}].

\bibitem{Bergstrom:2009fa}
L.~Bergstrom, J.~Edsjo, and G.~Zaharijas, {\it {Dark matter interpretation of
  recent electron and positron data}},  {\em Phys. Rev. Lett.} {\bf 103} (2009)
  031103, [\href{http://xxx.lanl.gov/abs/0905.0333}{{\tt arXiv:0905.0333}}].

\bibitem{Springel:2008by}
V.~Springel {\em et.~al.}, {\it {A blueprint for detecting supersymmetric dark
  matter in the Galactic halo}},  \href{http://xxx.lanl.gov/abs/0809.0894}{{\tt
  arXiv:0809.0894}}.

\bibitem{Diemand:2008in}
J.~Diemand {\em et.~al.}, {\it {Clumps and streams in the local dark matter
  distribution}},  {\em Nature} {\bf 454} (2008) 735--738,
  [\href{http://xxx.lanl.gov/abs/0805.1244}{{\tt arXiv:0805.1244}}].

\bibitem{Sommerfeld}
A.~Sommerfeld {\em Annalen der Physik} {\bf 403} (1931) 257.

\bibitem{Hisano:2002fk}
J.~Hisano, S.~Matsumoto, and M.~M. Nojiri, {\it {Unitarity and higher-order
  corrections in neutralino dark matter annihilation into two photons}},  {\em
  Phys. Rev.} {\bf D67} (2003) 075014,
  [\href{http://xxx.lanl.gov/abs/hep-ph/0212022}{{\tt hep-ph/0212022}}].

\bibitem{Hisano:2003ec}
J.~Hisano, S.~Matsumoto, and M.~M. Nojiri, {\it {Explosive dark matter
  annihilation}},  {\em Phys. Rev. Lett.} {\bf 92} (2004) 031303,
  [\href{http://xxx.lanl.gov/abs/hep-ph/0307216}{{\tt hep-ph/0307216}}].

\bibitem{Cirelli:2007xd}
M.~Cirelli, A.~Strumia, and M.~Tamburini, {\it {Cosmology and Astrophysics of
  Minimal Dark Matter}},  {\em Nucl. Phys.} {\bf B787} (2007) 152--175,
  [\href{http://xxx.lanl.gov/abs/0706.4071}{{\tt arXiv:0706.4071}}].

\bibitem{ArkaniHamed:2008qn}
N.~Arkani-Hamed, D.~P. Finkbeiner, T.~R. Slatyer, and N.~Weiner, {\it {A Theory
  of Dark Matter}},  {\em Phys. Rev.} {\bf D79} (2009) 015014,
  [\href{http://xxx.lanl.gov/abs/0810.0713}{{\tt arXiv:0810.0713}}].

\bibitem{Pospelov:2008jd}
M.~Pospelov and A.~Ritz, {\it {Astrophysical Signatures of Secluded Dark
  Matter}},  {\em Phys. Lett.} {\bf B671} (2009) 391--397,
  [\href{http://xxx.lanl.gov/abs/0810.1502}{{\tt arXiv:0810.1502}}].

\bibitem{MarchRussell:2008tu}
J.~D. March-Russell and S.~M. West, {\it {WIMPonium and Boost Factors for
  Indirect Dark Matter Detection}},  {\em Phys. Lett.} {\bf B676} (2009)
  133--139, [\href{http://xxx.lanl.gov/abs/0812.0559}{{\tt arXiv:0812.0559}}].

\bibitem{Iengo:2009ni}
R.~Iengo, {\it {Sommerfeld enhancement: general results from field theory
  diagrams}},  {\em JHEP} {\bf 05} (2009) 024,
  [\href{http://xxx.lanl.gov/abs/0902.0688}{{\tt arXiv:0902.0688}}].

\bibitem{Cassel:2009wt}
S.~Cassel, {\it {Sommerfeld factor for arbitrary partial wave processes}},
  {\em J. Phys.} {\bf G37} (2010) 105009,
  [\href{http://xxx.lanl.gov/abs/0903.5307}{{\tt arXiv:0903.5307}}].

\bibitem{Feldman:2008xs}
D.~Feldman, Z.~Liu, and P.~Nath, {\it {PAMELA Positron Excess as a Signal from
  the Hidden Sector}},  {\em Phys. Rev.} {\bf D79} (2009) 063509,
  [\href{http://xxx.lanl.gov/abs/0810.5762}{{\tt arXiv:0810.5762}}].

\bibitem{Ibe:2008ye}
M.~Ibe, H.~Murayama, and T.~T. Yanagida, {\it {Breit-Wigner Enhancement of Dark
  Matter Annihilation}},  {\em Phys. Rev.} {\bf D79} (2009) 095009,
  [\href{http://xxx.lanl.gov/abs/0812.0072}{{\tt arXiv:0812.0072}}].

\bibitem{Guo:2009aj}
W.-L. Guo and Y.-L. Wu, {\it {Enhancement of Dark Matter Annihilation via
  Breit-Wigner Resonance}},  {\em Phys. Rev.} {\bf D79} (2009) 055012,
  [\href{http://xxx.lanl.gov/abs/0901.1450}{{\tt arXiv:0901.1450}}].

\bibitem{Fairbairn:2008fb}
M.~Fairbairn and J.~Zupan, {\it {Two component dark matter}},  {\em JCAP} {\bf
  0907} (2009) 001, [\href{http://xxx.lanl.gov/abs/0810.4147}{{\tt
  arXiv:0810.4147}}].

\bibitem{Feldman:2009wv}
D.~Feldman, Z.~Liu, P.~Nath, and B.~D. Nelson, {\it {Explaining PAMELA and WMAP
  data through Coannihilations in Extended SUGRA with Collider Implications}},
  {\em Phys. Rev.} {\bf D80} (2009) 075001,
  [\href{http://xxx.lanl.gov/abs/0907.5392}{{\tt arXiv:0907.5392}}].

\bibitem{Boehm:2003ha}
C.~Boehm, P.~Fayet, and J.~Silk, {\it {Light and heavy dark matter particles}},
   {\em Phys. Rev.} {\bf D69} (2004) 101302,
  [\href{http://xxx.lanl.gov/abs/hep-ph/0311143}{{\tt hep-ph/0311143}}].

\bibitem{Hur:2007ur}
T.~Hur, H.-S. Lee, and S.~Nasri, {\it {A Supersymmetric U(1) -prime model with
  multiple dark matters}},  {\em Phys. Rev.} {\bf D77} (2008) 015008,
  [\href{http://xxx.lanl.gov/abs/0710.2653}{{\tt arXiv:0710.2653}}].

\bibitem{Adibzadeh:2008pe}
M.~Adibzadeh and P.~Q. Hung, {\it {The relic density of shadow dark matter
  candidates}},  {\em Nucl. Phys.} {\bf B804} (2008) 223--249,
  [\href{http://xxx.lanl.gov/abs/0801.4895}{{\tt arXiv:0801.4895}}].

\bibitem{Feng:2008ya}
J.~L. Feng and J.~Kumar, {\it {The WIMPless Miracle: Dark-Matter Particles
  without Weak- Scale Masses or Weak Interactions}},  {\em Phys. Rev. Lett.}
  {\bf 101} (2008) 231301, [\href{http://xxx.lanl.gov/abs/0803.4196}{{\tt
  arXiv:0803.4196}}].

\bibitem{Zurek:2008qg}
K.~M. Zurek, {\it {Multi-Component Dark Matter}},  {\em Phys. Rev.} {\bf D79}
  (2009) 115002, [\href{http://xxx.lanl.gov/abs/0811.4429}{{\tt
  arXiv:0811.4429}}].

\bibitem{Batell:2009vb}
B.~Batell, M.~Pospelov, and A.~Ritz, {\it {Direct Detection of Multi-component
  Secluded WIMPs}},  {\em Phys. Rev.} {\bf D79} (2009) 115019,
  [\href{http://xxx.lanl.gov/abs/0903.3396}{{\tt arXiv:0903.3396}}].

\bibitem{Profumo:2009tb}
S.~Profumo, K.~Sigurdson, and L.~Ubaldi, {\it {Can we discover multi-component
  WIMP dark matter?}},  {\em JCAP} {\bf 0912} (2009) 016,
  [\href{http://xxx.lanl.gov/abs/0907.4374}{{\tt arXiv:0907.4374}}].

\bibitem{Zhang:2009dd}
H.~Zhang, C.~S. Li, Q.-H. Cao, and Z.~Li, {\it {A Dark Matter Model with
  Non-Abelian Gauge Symmetry}},  {\em Phys. Rev.} {\bf D82} (2010) 075003,
  [\href{http://xxx.lanl.gov/abs/0910.2831}{{\tt arXiv:0910.2831}}].

\bibitem{Gao:2010pg}
X.~Gao, Z.~Kang, and T.~Li, {\it {The Supersymmetric Standard Models with Decay
  and Stable Dark Matters}},  {\em Eur. Phys. J.} {\bf C69} (2010) 467--480,
  [\href{http://xxx.lanl.gov/abs/1001.3278}{{\tt arXiv:1001.3278}}].

\bibitem{Feldman:2010wy}
D.~Feldman, Z.~Liu, P.~Nath, and G.~Peim, {\it {Multicomponent Dark Matter in
  Supersymmetric Hidden Sector Extensions}},  {\em Phys. Rev.} {\bf D81} (2010)
  095017, [\href{http://xxx.lanl.gov/abs/1004.0649}{{\tt arXiv:1004.0649}}].

\bibitem{Gogoladze:2009gi}
I.~Gogoladze, N.~Okada, and Q.~Shafi, {\it {Type II Seesaw and the PAMELA/ATIC
  Signals}},  {\em Phys. Lett.} {\bf B679} (2009) 237--241,
  [\href{http://xxx.lanl.gov/abs/0904.2201}{{\tt arXiv:0904.2201}}].

\bibitem{Guo:2010vy}
W.-L. Guo, Y.-L. Wu, and Y.-F. Zhou, {\it {Exploration of decaying dark matter
  in a left-right symmetric model}},  {\em Phys. Rev.} {\bf D81} (2010) 075014,
  [\href{http://xxx.lanl.gov/abs/1001.0307}{{\tt arXiv:1001.0307}}].

\bibitem{Zavala:2009mi}
J.~Zavala, M.~Vogelsberger, and S.~D.~M. White, {\it {Relic density and CMB
  constraints on dark matter annihilation with Sommerfeld enhancement}},  {\em
  Phys. Rev.} {\bf D81} (2010) 083502,
  [\href{http://xxx.lanl.gov/abs/0910.5221}{{\tt arXiv:0910.5221}}].

\bibitem{Hannestad:2010zt}
S.~Hannestad and T.~Tram, {\it {Sommerfeld Enhancement of DM Annihilation:
  Resonance Structure, Freeze-Out and CMB Spectral Bound}},  {\em JCAP} {\bf
  1101} (2011) 016, [\href{http://xxx.lanl.gov/abs/1008.1511}{{\tt
  arXiv:1008.1511}}].

\bibitem{Finkbeiner:2010sm}
D.~P. Finkbeiner, L.~Goodenough, T.~R. Slatyer, M.~Vogelsberger, and N.~Weiner,
  {\it {Consistent Scenarios for Cosmic-Ray Excesses from Sommerfeld-Enhanced
  Dark Matter Annihilation}},  \href{http://xxx.lanl.gov/abs/1011.3082}{{\tt
  arXiv:1011.3082}}.

\bibitem{Lattanzi:2008qa}
M.~Lattanzi and J.~I. Silk, {\it {Can the WIMP annihilation boost factor be
  boosted by the Sommerfeld enhancement?}},  {\em Phys. Rev.} {\bf D79} (2009)
  083523, [\href{http://xxx.lanl.gov/abs/0812.0360}{{\tt arXiv:0812.0360}}].

\bibitem{Kamionkowski:2008gj}
M.~Kamionkowski and S.~Profumo, {\it {Early Annihilation and Diffuse
  Backgrounds in Models of Weakly Interacting Massive Particles in Which the
  Cross Section for Pair Annihilation Is Enhanced by 1/v}},  {\em Phys. Rev.
  Lett.} {\bf 101} (2008) 261301,
  [\href{http://xxx.lanl.gov/abs/0810.3233}{{\tt arXiv:0810.3233}}].

\bibitem{Feng:2009mn}
J.~L. Feng, M.~Kaplinghat, H.~Tu, and H.-B. Yu, {\it {Hidden Charged Dark
  Matter}},  {\em JCAP} {\bf 0907} (2009) 004,
  [\href{http://xxx.lanl.gov/abs/0905.3039}{{\tt arXiv:0905.3039}}].

\bibitem{Feng:2009hw}
J.~L. Feng, M.~Kaplinghat, and H.-B. Yu, {\it {Halo Shape and Relic Density
  Exclusions of Sommerfeld- Enhanced Dark Matter Explanations of Cosmic Ray
  Excesses}},  {\em Phys. Rev. Lett.} {\bf 104} (2010) 151301,
  [\href{http://xxx.lanl.gov/abs/0911.0422}{{\tt arXiv:0911.0422}}].

\bibitem{Feng:2010zp}
J.~L. Feng, M.~Kaplinghat, and H.-B. Yu, {\it {Sommerfeld Enhancements for
  Thermal Relic Dark Matter}},  {\em Phys. Rev.} {\bf D82} (2010) 083525,
  [\href{http://xxx.lanl.gov/abs/1005.4678}{{\tt arXiv:1005.4678}}].

\end{thebibliography}

\providecommand{\href}[2]{#2}\begingroup\raggedright\endgroup

\end{document}